\documentclass[12pt]{article}
\usepackage[dvips]{color}
\usepackage{amsmath}
\usepackage{graphicx}
\usepackage{subfigure}
\usepackage{epsfig}
\usepackage{amsmath}
\usepackage{cite}
\usepackage{color}
\usepackage{subfigure}
\usepackage{multirow}

\begin{document}
\begin{center}
\Large{\bf   Exploring the Phase Transition in Charged Gauss-Bonnet Black Holes: A Holographic Thermodynamics Perspectives}\\
 \small \vspace{1cm}
 {\bf Jafar Sadeghi $^{\star}$\footnote {Email:~~~pouriya@ipm.ir}}, \quad
 {\bf Mohammad Reza Alipour $^{\star}$\footnote {Email:~~~mr.alipour@stu.umz.ac.ir}}\quad
 {\bf Mohammad Ali S. Afshar $^{\star}$\footnote {Email:~~~m.a.s.afshar@gmail.com}}\quad\\\vspace{0.2cm}
 {\bf Saeed Noori Gashti$^{\dag,\star}$\footnote {Email:~~~saeed.noorigashti@stu.umz.ac.ir}}, \quad\\
\vspace{0.5cm}$^{\star}${Department of Physics, Faculty of Basic
Sciences,\\
University of Mazandaran
P. O. Box 47416-95447, Babolsar, Iran}\\
\vspace{0.5cm}$^{\dag}${School of Physics, Damghan University, P. O. Box 3671641167, Damghan, Iran}
\small \vspace{1cm}
\end{center}
\begin{abstract}
In this paper, we delve into the study of thermodynamics and phase transition of charged Gauss-Bonnet black holes within the context of anti-de Sitter (AdS) space, with particular emphasis on the central charge's role within the dual conformal field theory (CFT). We employ a holographic methodology that interprets the cosmological constant and the Newton constant as thermodynamic variables, leading to the derivation of a modified first law of thermodynamics that incorporates the thermodynamic volume and pressure. Our findings reveal that the central charge of the CFT is intrinsically linked to the variation of these constants, and its stability can be ensured by simultaneous adjustment of these constants. We further explore the phase structures of the black holes, utilizing the free energy. Our research uncovers the existence of a critical value of the central charge, beyond which the phase diagram displays a first-order phase transition between small and large black holes. We also delve into the implications of our findings on the complexity of the CFT. Our conclusions underscore the significant role of the central charge in the holographic thermodynamics and phase transition of charged Gauss-Bonnet black holes. Furthermore, we conclude that while the central charge considered provides suitable and satisfactory solutions for this black hole in 4 and 5 dimensions, it becomes necessary to introduce a unique central charge for this structure of modified gravity. In essence, the central charge in holographic thermodynamics is not a universal value and requires modification in accordance with different modified gravities. Consequently, the physics of the problem will significantly deviate from the one discussed in this article, indicating a rich and complex landscape for future work.
\\
Keywords: Holographic Thermodynamics, Phase Transitions, Central Charge\\\\
\end{abstract}
\tableofcontents
\section{Introduction}
Without a doubt, black holes function as essential tools in the investigation and progression of quantum gravity, providing empirical evidence and data that illuminate this elusive theory.
While numerous branches of physics contribute to the behavioral study of various black hole models and their theoretical development, the role of thermodynamics has become increasingly prominent\cite{10000,1000,1001,1002,1003,1004,1005}. The significance of thermodynamics is primarily attributed to its parameters, which frequently appear in the form of energy, leading to diminished sensitivity in relativistic calculations which this attribute is crucial when examining a system within a relativistic framework. Moreover, thermodynamics has proven to be successful in clarifying black hole processes, thus inspiring a growing number of scientists to employ this branch\cite{10000,1000,1001,1002,1003,1004,1005}. Currently, black hole thermodynamics is acknowledged as a separate and dependable branch within the sphere of modern cosmological physics\cite{1006,1007,1008,1009,1010,1011,1011',1011''}.
A critical event in the history of black hole thermodynamics was the incorporation of the cosmological constant $\Lambda$, as a thermodynamic parameter. This parameter took on the role of pressure in the first law of thermodynamics, establishing the basis for a new viewpoint on black hole thermodynamics, known as black hole chemistry \cite{1,2,3,4,5,6,7,8,9}.\\\\
Phase transition is a phenomenon in which a system changes its state or properties abruptly under certain conditions, which is often studied in the form of first or second order of phase transition. The key difference between the two is that first-order transitions involve a discontinuity in the first derivative of the free energy with respect to some thermodynamic variable (like temperature), while second-order transitions involve a discontinuity in the second derivative of the free energy\cite{1013,1014}.
Black holes can also undergo phase transitions, and phase transitions in black holes are a captivating subject in theoretical physics, often explored within the framework of the AdS/CFT correspondence\cite{1015,1016,1017}. First-order phase transitions in black holes can be conceptualized as a transformation from one type of black hole to another, with this transition involving the release or absorption of energy. This is akin to the energy exchange that occurs during the phase transition of water from a liquid to a gas state. Second-order phase transitions in black holes, conversely, are more nuanced and involve changes in the properties of the black hole as some parameter is varied. These transitions do not involve a latent heat\cite{1018,1019}. An example of a second-order phase transition in black holes is the Hawking-Page phase transition\cite{1020,1021,1022,1023,1024}. This is a phase transition between a large black hole in anti-de Sitter (AdS) space and thermal AdS space without a black hole. At low temperatures, the thermal AdS space is more stable, while at high temperatures, the black hole phase is more stable. The transition between these two phases is a second-order phase transition\cite{1020,1021,1022,1023,1024}.
It should be noted that the study of phase transition in black holes can be investigated with various methods,
among which we can refer to the usual classical methods, through the examination of the continuity of the derivatives of thermodynamic parameters and its stabilities, or the method of thermodynamic topology which each has its own characteristics. For example, the topological method can reveal the deep connection between black hole physics and ordinary thermodynamics, as well as the quantum gravity effects in the bulk. It can also provide new insights into the holographic principle\cite{a,b,c,d,e,f,g,h,i,j,k,l,m,n}.\\\\
With the introduction of the holographic principle, which conjectures that the information content of a region of space can be encoded on a lower dimension of it, a new idea was opened for scientists under the name of holographic black hole chemistry\cite{1025,1026,1027,1028}.
The holographic principle implies that there is a duality between gravity in the bulk and quantum field theory on the boundary, known as the AdS/CFT correspondence\cite{1029,1030,1031,1032}. This duality allows us to study the thermodynamics of black holes in Anti-de Sitter (AdS) space using the properties of a conformal field theory (CFT) at finite temperatures on the boundary. In particular, one can include the cosmological constant as a dynamical variable and interpret it as a thermodynamic pressure in the bulk and a scale parameter in the boundary. This leads to a new perspective on black hole thermodynamics, called black hole chemistry, where one can define new quantities such as enthalpy, Gibbs free energy, and chemical potential for black holes\cite{1025,1026,1027,1028}. Holographic black hole chemistry aims to understand the properties and phase transitions of black holes in AdS space using holographic duality. For example, one can study the Hawking-Page transition, which is a phase transition between thermal AdS space and a large AdS black hole, from the point of view of the CFT undergoing a confinement/deconfinement transition. One can also study more general phase transitions involving charged or rotating black holes, or higher-order curvature corrections\cite{1025,1026,1027,1028}..
Another interesting result is that black holes in AdS space obey a generalized Smarr relation, which is an identity relating to the mass, area, charge, angular momentum, and pressure of the black hole.
The holographic interpretation of black hole thermodynamics also faces some challenges. For instance, the first law of black hole thermodynamics could not be directly related to the thermodynamics of the holographic dual CFT \cite{10,11,12,13,14}.\\\\
With these interpretations, what is important for us and the motivation of our work is that the variation of the cosmological constant $\Lambda$ is related to both the variation of the central charge $C$ and the CFT volume $V$ of the boundary CFT. This implies that the law of thermodynamics needs to be modified when we consider the volume and its conjugate, i.e., the thermodynamic pressure in the bulk. Moreover, variations of the fundamental constants, such as the Newton constant and the cosmological constant, suggest that more fundamental theories exist \cite{15,16,17}. Cong et al. solved this problem by showing that changing Newton's constant $G$ together with the cosmological constant $\Lambda$ can keep the central charge $C$ of the dual CFT constant. Therefore, the first law can be written in a mixed form that includes the appropriate new thermodynamic volume and chemical potential. This establishes a connection between holographic and bulk thermodynamics by using the duality relation of the central charge. The research on free energy reveals the phase structures with a critical value of the central charge, above which the phase diagram exhibits a swallowtail behavior that indicates a first-order phase transition. Following the same approach, Newton's constant $G$ can be treated as a variable, so that the central charge can be introduced into the first law and the thermodynamics can be discussed for different black holes \cite{18,19,20,21,21',22,23,24,25,26,27,28}. In other words, it can be said that one of the remarkable results of black hole chemistry is that AdS black holes undergo phase transitions that are fully analogous to those of ordinary thermodynamic systems.\\\\
However, the holographic interpretation of extended thermodynamics was unclear for a long time. The first attempts suggested that, according to the AdS-CFT correspondence, the $V\delta P$ term should correspond to a $\mu\delta C$ term in the dual CFT, where $C$ is the central charge and $\mu$ is the thermodynamically conjugate chemical potential\cite{29,30,31,32,33,34,35,29',30',300',31',32',33'}.
This new perspective motivated us to study the effect of dimensions, gravity, and gauge corrections according to these new changes on the thermodynamics of the boundary.
For this reason, to investigate the thermodynamic behavior and phase transition of a black hole under this form of equations, we chose the special Gauss-Bonet black hole and then compared the obtained results with other studies.\cite{18}.
So, based on the above explanations, we arrange the article as follows:\\
In section 2, we provide a brief overview of the thermodynamics of Gauss-Bonnet-AdS black holes in 4 and 5 dimensions. In section 3, we investigate the generalized mass/energy formulas for these black holes, such as the extended, mixed, and CFT thermodynamics. In section 4, we discuss the results of our calculations; finally, in section 5, we present our conclusion.
\section{Thermodynamics of black holes}
Thermodynamics of black holes is the area of study that seeks to reconcile the laws of thermodynamics with the existence of black hole event horizons. It involves the concepts of black hole entropy, temperature, radiation, and phase transitions. It also has implications for quantum gravity and the holographic principle. Thermodynamics is the branch of physics that deals with the relationships between heat, work, energy, and entropy. Thermodynamics can be applied to various systems, including black holes, to study their properties and behavior. For example, one can define the temperature, pressure, and chemical potential of a black hole, and derive the first law of thermodynamics for it. One can also study the phase transitions and critical phenomena of black holes, such as the Hawking-Page transition, which is analogous to the liquid-gas transition in ordinary thermodynamics.\\

The holographic principle and thermodynamics are related in several ways. One way is that the holographic principle can be derived from the laws of thermodynamics applied to the event horizon of a black hole. Another way is that the holographic principle can be used to map a gravitational theory in a higher-dimensional space, such as anti-de Sitter space (AdS), to a quantum field theory in a lower-dimensional space, such as the conformal field theory (CFT) on the boundary of AdS. This correspondence, known as the AdS/CFT correspondence, allows one to use the tools of quantum field theory and thermodynamics to study the properties of quantum gravity and black holes. For example, one can calculate the entropy, free energy, and phase diagrams of black holes in AdS space using the CFT on the boundary. In this section, we review the thermodynamic quantities of AdS black hole with Gauss-Bonnet (GB) corrections. For AdS black holes, the thermodynamic properties are determined by the negative cosmological constant and its associated thermodynamic pressure\cite{3,b,18,1033,1034,33',33''},
\begin{equation}\label{eq1}
\Lambda=-\frac{(d-2)(d-1)}{2\ell^2},   \qquad     P=-\frac{\Lambda}{8\pi G},
\end{equation}
where $\ell$ represents the AdS radius of the d-dimensional space time.
\subsection{Gauss-Bonnet-AdS black hole}
We examine the Gauss-Bonnet-AdS black hole in two states $d\geq 5$ and $d=4$.
\subsubsection{$d\geq5$}
The action of the d-dimensional Einstein-Maxwell theory, incorporating the negative cosmological constant and Gauss-Bonnet term, is presented\cite{3,b,18,1033,1034,33',33''},
\begin{equation}\label{eq2}
S=\frac{1}{16\pi G}\int d^dx \sqrt{-g}\bigg[R-2\Lambda+\alpha_{GB}(R_{\mu\nu\sigma\rho}R^{\mu\nu\sigma\rho}+R^2-4R_{\mu\nu}R^{\mu\nu})-4\pi G F_{\mu\nu}F^{\mu\nu}\bigg].
\end{equation}
The equation involves the electromagnetic field tensor represented by $F_{\mu\nu}=\partial_{\mu}A_{\nu}-\partial_{\nu}A_{\mu}$, the Gauss-Bonnet coupling constant $\alpha_{GB}$, the Newton's constant $G$, and the cosmological constant represented by $\Lambda$. The Gauss-Bonnet coupling constant has a positive value in string theory and has a dimension of $(length)^2$.
We are focusing on the spherical topology of the horizon, meaning that $k = 1$. Therefore, the metric can be expressed in the following form\cite{3,b,18,1033,1034,33',33''},
\begin{equation}\label{eq3}
ds^2=-f(r)dt^2+f^{-1}(r) dr^2+r^2(d\theta^2+\sin^2\theta d\phi^2+\cos^2\theta d\Omega^2_{d-4}),
\end{equation}
 and,
 \begin{equation}\label{eq4}
f(r)=1+\frac{r^2}{2\beta}\big[1-\bigg(1+\frac{64\pi \beta GM}{(d-2)\Sigma r^{d-1}}-\frac{8G\beta Q^2 }{(d-2)(d-3)r^{2(d-2)}}-\frac{4\beta}{\ell^2}\bigg)^{\frac{1}{2}}\big],
\end{equation}
where $\Sigma$, $Q$ and $M$ are the area of
$(d - 2)$-dimensional unit sphere, electric charge, and the mass of the black hole respectively, also $\beta= (d-4)(d-3)\alpha_{GB}$. By solving $f(r_+)=0$ and using the $T=\frac{f^{\prime}(r_+)}{4\pi}$ relation, we can obtain the mass and temperature of the black hole\cite{3,b,18,1033,1034,33',33''},
 \begin{equation}\label{eq5}
M=\frac{\Sigma (d-2) r_+^{d-3}}{16\pi G r_+^2 \ell^2}(r_+^2 \ell^2+\beta \ell^2 + r_+^4)+\frac{\Sigma Q^2}{8\pi (d-3)r_+^{d-3}},
\end{equation}
and,
 \begin{equation}\label{eq6}
T=\frac{(d-1)(d-2)r_+^{2(d-2)}+(d-2)(d-3)\ell^2 r_+^{2(d-3)}+(d-5)(d-2)\beta r_+^{2(d-4)}-2GQ^2}{4\pi r_+(r_+^2+2\beta)\ell^2(d-2)r_+^{2(d-4)}}.
\end{equation}
Additionally, the black hole's entropy can be expressed as\cite{18},
 \begin{equation}\label{eq7}
S=\frac{\Sigma r_+^{d-2}}{4G}\bigg(1+\frac{2(d-2)\beta}{(d-4)r_+^2} \bigg).
\end{equation}
\subsubsection{$d=4$}
Note that although at d=4, the GB coupling is topological, there have been some recent efforts in the literature to constrain such coupling constant using both theoretical and observational techniques\cite{1033,1034}. It is a widely accepted fact that within the context of GB gravity, black hole solutions with static and spherically symmetric properties exist when $d\geq 5$. In four-dimensional spacetime, the GB term does not contribute to the field equation and therefore there are no GB black holes in this scenario. However, Glavan and Lin\cite{1035} made an interesting discovery by rescaling the GB coupling parameter $\alpha \rightarrow \alpha/(d-4)$ and taking the limit $d = 4$, which resulted in a non-trivial black hole solution in four dimensions. This solution was later extended to include charged cases in an AdS space. If we take the limit $d \rightarrow 4$ into consideration, we can find the solution for this black hole\cite{3,b,18,1033,1034,33',33''},
 \begin{equation}\label{eq8}
ds^2=-f(r) dt^2+f^{-1}(r) dr^2+ r^2(d\theta^2 +\sin^2 \theta d\phi^2),
\end{equation}
where the metric function is,
 \begin{equation}\label{eq9}
f(r)=1+\frac{r^2}{2\alpha}\bigg[1-\bigg(1-4\alpha(\frac{GQ^2}{r^4}+\frac{1}{\ell^2}-\frac{2MG}{r^3}) \bigg)^{1/2} \bigg].
\end{equation}
By solving $f(r_+)=0$, we can get the mass of the black hole\cite{3,b,18,1033,1034,33',33''},
 \begin{equation}\label{eq10}
M=\frac{G l^2 Q^2+\alpha  l^2+l^2 r_+^2+r_+^4}{2 G l^2 r_+}.
\end{equation}
Also, thermodynamic quantities such as entropy and temperature are obtained through the following relationship\cite{18},
 \begin{equation}\label{eq11}
T=\frac{-G l^2 Q^2-\alpha  l^2+l^2 r_+^2+3 r_+^4}{4 \pi  l^2 \left(r_+^3+2 \alpha  r_+\right)},
\end{equation}
 \begin{equation}\label{eq12}
S=\frac{\pi  r_+^2+4 \pi  \alpha  \log \left(\frac{r_+}{\sqrt{\alpha }}\right)}{G}.
\end{equation}
According to equation 12 and also the positive entropy, we have $\alpha>0$
\section{Generalized mass/energy formulas}
In this section, we try to introduce the thermodynamics of black holes in different spaces. In each of these spaces, the energy of black holes can be a different interpretation of the mass of black holes. Therefore, we investigate the first law of thermodynamics and the Smarr relation for black holes considered in extended thermodynamics, mixed thermodynamics, and CFT thermodynamics.
We also calculate thermodynamic quantities for black holes considered in different spaces.
\subsection{Extended thermodynamics}
For black holes that are in AdS spacetime, we can consider a negative cosmological constant proportional to the pressure. In this case, its conjugate quantity is the volume of the black hole.
In the following, we state the first law of thermodynamics for black holes considered in extended space. Note that in this thermodynamics, the mass of the black hole plays the role of the system's enthalpy.
\subsubsection{GB-AdS black hole}
Since the first law of thermodynamics and Smarr's relation are the same for black holes in $d=4$ and $d\geq5$ dimensions, we investigate the GB-AdS black hole in $d\geq 4$ mode in the extended space (with the condition that $d=4$, we have $\beta=\alpha$ and $d\geq5$, we have $\beta=(d-3)(d-4))\alpha$. The black hole's first law and Smarr formula in extended thermodynamics can be expressed as follows\cite{3,b,18,1033,1034,33',33''},
\begin{equation}\label{eq19}
\delta M=T \delta S+ \Phi \delta Q+ V \delta P+ \mathcal{A} \delta\beta,
\end{equation}
\begin{equation}\label{eq20}
M= \frac{d-2}{d-3}T S+ \Phi Q-\frac{2}{d-3} V P+ \frac{2}{d-3} \mathcal{A} \beta,
\end{equation}
where $\mathcal{A}$ in the extended space is the conjugate of $\beta$.
\subsection{Mixed thermodynamics}
We aim to convert the bulk thermodynamic first law into a format that includes the boundary central charge. To do this, we will use the holographic dual relationship between the $AdS$ scale $\ell$, the central charge $C$, and Newton's constant $G$ in Einstein's gravity,
\begin{equation}\label{eq23}
C=\frac{k \ell^{d-2}}{16 \pi G}.
\end{equation}
The value of the $k$ factor is influenced by the specific characteristics of the system at the boundary.
Also, by using equations \eqref{eq1} and \eqref{eq23}, we can express Newton's constant in terms of bulk thermodynamic pressure and boundary central charge,
\begin{equation}\label{eq24}
G=\bigg(\frac{k}{16\pi C}\bigg)^{\frac{2}{d}}\bigg(\frac{(d-2)(d-1)}{16\pi P}  \bigg)^{\frac{d-2}{d}}.
\end{equation}
So, we can rewrite the first law of thermodynamics for the black holes under consideration.
\subsubsection{GB-AdS black hole}
The mixed form of the first law of thermodynamics for a GB-AdS black hole, expressed in a combination of bulk and boundary conditions, is as follows,
\begin{equation}\label{eq25}
\delta M=T \delta S+ \Phi \delta Q+ V_{bb} \delta P+\mu_{bb} \delta C+ \mathcal{A} \delta \beta,
\end{equation}
The chemical potential and thermodynamic volume have been defined as,
\begin{equation}\label{eq26}
\mu_{bb}=\frac{2P(V_{bb}-V)}{(d-2) C}     , \qquad    V_{bb}= \frac{2M+(d-4)\Phi Q+4\beta \mathcal{A}}{2Pd},
\end{equation}
In addition, the mass of the black hole can be expressed in terms of the central charge by using equations \eqref{eq1}, \eqref{eq5}, and \eqref{eq23}. This can be written as follows,
\begin{equation}\label{eq27}
M=\frac{P \Sigma  r_+^{d-5} \left(\frac{k}{C}\right)^{-2/d} \left(r_+^4 \left(\frac{k}{C}\right)^{2/d}+\beta  \left(\frac{(d-2) (d-1)}{P}\right)^{2/d}+r_+^2 \left(\frac{(d-2) (d-1)}{P}\right)^{2/d}\right)}{d-1}+\frac{Q^2 \Sigma  r_+^{3-d}}{8 \pi  (d-3)}.
\end{equation}
Note that, according to equations \eqref{eq7} and \eqref{eq12}, we cannot rewrite the event horizon $(r_+)$ in terms of entropy $(S)$. As a result, we obtain thermodynamic quantities in terms of the event horizon. Therefore, according to equations \eqref{eq25} and \eqref{eq27}, the thermodynamic parameters $T, \Phi, V_{bb}, \mu_{bb}, \mathcal{A}$ are calculated as follows,
\begin{equation}\label{eq28}
\begin{split}
&T=\bigg(\frac{\partial M}{ \partial S} \bigg)_{Q,P,C,\beta}=\bigg(\frac{\partial M}{ \partial r_+}\bigg) \bigg(\frac{\partial r_+}{ \partial S}\bigg)= \frac{\left(\frac{(d-2) (d-1)}{P}\right)^{1-\frac{2}{d}} r_+^{5-d} \left(\frac{k}{C}\right)^{2/d}}{4 \pi  (d-2) \left(2 \beta +r_+^2\right)}\\
&\times \bigg[\frac{(3-d) Q^2 r_+^{2-d}}{8 \pi  (d-3)}+P r_+^{d-6} \left(\frac{\left(\frac{(d-2) (d-1)}{P}\right)^{2/d} \left(\frac{k}{C}\right)^{-2/d} \left(\beta  (d-5)+(d-3) r_+^2\right)}{d-1}+r_+^4\right) \bigg],
\end{split}
\end{equation}

\begin{equation}\label{eq29}
\begin{split}
\Phi=\bigg(\frac{\partial M}{ \partial Q} \bigg)_{S,P,C,\beta}= \frac{Q \Sigma  r_+^{3-d}}{4 \pi  (d-3)},
\end{split}
\end{equation}

\begin{equation}\label{eq30}
\begin{split}
&V_{bb}=\bigg(\frac{\partial M}{ \partial P} \bigg)_{S,Q,C,\beta}= \frac{\Sigma  r_+^{d-5} \left((d-2) \left(\frac{(d-2) (d-1)}{P}\right)^{2/d} \left(\beta +r_+^2\right) \left(\frac{k}{C}\right)^{-2/d}+d r_+^4\right)}{d (d-1)},
\end{split}
\end{equation}

\begin{equation}\label{eq31}
\begin{split}
&\mu_{bb}=\bigg(\frac{\partial M}{ \partial C} \bigg)_{S,Q,P,\beta}= -\frac{2 P \Sigma  r_+^{d-1}}{C (d-1) d}\\
&+\frac{2 k P \Sigma  r_+^{d-5} \left(\frac{k}{C}\right)^{-\frac{2}{d}-1} \left(r_+^4 \left(\frac{k}{C}\right)^{2/d}+\beta  \left(\frac{(d-2) (d-1)}{P}\right)^{2/d}+r_+^2 \left(\frac{(d-2) (d-1)}{P}\right)^{2/d}\right)}{C^2 (d-1) d}
\end{split}
\end{equation}

\begin{equation}\label{eq32}
\begin{split}
\mathcal{A}=\bigg(\frac{\partial M}{ \partial \beta} \bigg)_{S,Q,P,C}=\frac{P \Sigma  \left(\frac{(d-2). (d-1)}{P}\right)^{2/d} r_+^{d-5} \left(\frac{k}{C}\right)^{-2/d}}{d-1}
\end{split}
\end{equation}
For further study, you can refer to\cite{33'}, which investigated the thermodynamics of the GB-AdS black hole in the bulk-boundary space.
\subsection{CFT thermodynamics}
To determine the CFT thermodynamics of the charged AdS black hole, we need to rely on additional holographic connections between the quantities in the bulk and on the boundary.
In this case, we assume that the curvature radius of the boundary is denoted as $R$, which is distinct from the AdS radius $\ell$ in the bulk. The metric of the CFT, which exhibits conformal scaling invariance, can be expressed as follows,
\begin{equation}\label{eq41}
\begin{split}
ds^2=\omega^2(-dt^2+\ell^2 d\Omega_{d-2}^2).
\end{split}
\end{equation}
The dimensionless conformal factor, denoted by $\omega$, is allowed to vary freely, representing the conformal symmetry of the boundary theory. In the context of a spherical case, $d\Omega_{d-2}^2$ represents the metric on a ($d-2$)-dimensional sphere with a volume denoted by $\Omega_{d-2}=k$. We assume that $\omega$ does not depend on the boundary coordinates. In this case, the volume of the conformal field theory (CFT) is given by,
\begin{equation}\label{eq42}
\begin{split}
\mathcal{V}=k R^{d-2},
\end{split}
\end{equation}
where $R = \omega \ell$ represents the variable curvature radius of the manifold where the CFT exists. The change of the CFT volume $\mathcal{V}$ is therefore clearly unaffected by the change of the central charge $C$, which, in the context of Einstein's gravity, has a dual relationship, which is shown in equation \eqref{eq23} of this duality. Note that here we keep Newton's constant $G$ fixed and allow the bulk radius of curvature $\ell$ to change, which also changes the central charge $C$ according to equation \eqref{eq23}.
The holographic dictionary allows us to relate the bulk quantities $M, T, S, \Phi, Q$ to their corresponding boundary CFT counterparts $E, \tilde{T} , \tilde{S}, \tilde{\Phi} , \tilde{Q}$,
\begin{equation}\label{eq43}
\begin{split}
E=\frac{M}{\omega}, \qquad  \tilde{T}=\frac{T}{\omega}, \qquad  \tilde{S}=S, \qquad \tilde{Q}=\frac{Q \ell}{\sqrt{G}}, \qquad  \tilde{\Phi}=\frac{\Phi \sqrt{G}}{\omega \ell}.
\end{split}
\end{equation}
According to the relations above, we can rewrite the first law of thermodynamics for the charged AdS black hole at the CFT as\cite{29},
\begin{equation}\label{eq44}
\begin{split}
\delta E= \tilde{\Phi} \delta \tilde{Q}+ \mu \delta C-p\delta \mathcal{V} +\tilde{T} \delta S,
\end{split}
\end{equation}
where\cite{29}
\begin{equation}\label{eq45}
\begin{split}
\mu=\frac{E-\tilde{T}S-\tilde{\Phi} \tilde{Q}}{C},
\end{split}
\end{equation}
\begin{equation}\label{eq46}
\begin{split}
p=\frac{E}{(d-2)\mathcal{V}}.
\end{split}
\end{equation}
Smarr relation for the RN AdS black hole in obtained as follows,
\begin{equation}\label{eq299}
\begin{split}
E=\tilde{T} S+\tilde{\Phi} \tilde{Q}+\mu C
\end{split}
\end{equation}
Also, according to equation \eqref{eq44}, we find that $\mathcal{V}$ and $C$ change independently.
The aim of this paper is to investigate the implications of this proposal for the CFT thermodynamics of $GB$ AdS black holes.
\subsubsection{GB-AdS black hole}
In this section, we attempt to derive the first law of thermodynamics in the CFT mode for the GB-AdS black hole and utilize it to establish the correspondence between bulk and boundary quantities.
Here, using the scale transformation $E=\frac{M}{\omega}$ (where $\omega=\frac{R}{\ell}$) and equations \eqref{eq19}, \eqref{eq20} and \eqref{eq23}, we have,
\begin{equation}\label{eq47}
\begin{split}
& \delta\bigg( \frac{M}{\omega}\bigg)= \frac{T}{\omega} \delta \bigg(\frac{A}{4G} \bigg)+\bigg(\frac{M}{\omega}-\frac{TS}{\omega}-\frac{Q\Phi}{\omega}-\frac{A\beta}{\omega} \bigg) \frac{\delta (k \ell^{d-2}/G)}{k \ell^{d-2}/G}\\
&-\frac{M}{\omega(d-2)}\frac{\delta (kR^{d-2})}{kR^{d-2}}+\frac{\Phi \sqrt{G}}{\omega\ell} \delta \bigg(\frac{Q\ell}{\sqrt{G}} \bigg)+\frac{A}{\omega \ell} \bigg(\ell \delta \beta+(d-4)\beta \delta \ell \bigg).
\end{split}
\end{equation}
\emph{ Therefore, according to the last part of relation \eqref{eq47}, we can only have the first law of CFT when $d=4$ or $d=5$.} This result is very interesting because the first law of CFT does not hold in $d$ dimensions for the GB-AdS black hole. In this case, we have,
\begin{equation}\label{eq48}
\begin{split}
&  d=4\qquad \rightarrow \qquad \delta\bigg( \frac{M}{\omega}\bigg)= \frac{T}{\omega} \delta \bigg(\frac{A}{4G} \bigg)+\bigg(\frac{M}{\omega}-\frac{TS}{\omega}-\frac{Q\Phi}{\omega}-\frac{A\beta}{\omega} \bigg) \frac{\delta (k \ell^{2}/G)}{k \ell^{2}/G}\\
&-\frac{M}{2\omega}\frac{\delta (kR^{2})}{kR^{2}}+\frac{\Phi \sqrt{G}}{\omega\ell} \delta \bigg(\frac{Q\ell}{\sqrt{G}} \bigg)+\frac{\mathcal{A}}{\omega }\delta \beta.
\end{split}
\end{equation}
According to equation \eqref{eq48}, the dictionary between bulk and boundary quantities for the GB-AdS black hole is obtained as follows,
\begin{equation}\label{eq49}
\begin{split}
&  d=4\qquad \rightarrow \qquad \delta E= \tilde{T} \delta \tilde{S}+\mu \delta C-p \delta \mathcal{V}+ \tilde{\Phi} \delta \tilde{Q}+ \tilde{A} \delta\tilde{\beta}\\
& E=\frac{M}{\omega}, \qquad  \tilde{S}=S,\qquad \tilde{T}=\frac{T}{\omega}, \qquad  \tilde{\Phi}=\frac{\Phi\sqrt{G}}{\omega\ell},\qquad \tilde{Q}=\frac{Q\ell}{\sqrt{G}}, \qquad \tilde{\mathcal{A}}=\frac{\mathcal{A}}{\omega},\qquad \tilde{\beta}=\beta,
\end{split}
\end{equation}
where, using equations \eqref{eq48}, and \eqref{eq49}, we will have,
\begin{equation}\label{eq50}
\begin{split}
&  \mu=\frac{1}{C}(E-\tilde{T}S-\tilde{\Phi}\tilde{Q}-\tilde{\mathcal{A}}\beta)\\
&p=\frac{E}{2\mathcal{V}}, \qquad \mathcal{V}=k R^2.
\end{split}
\end{equation}
Therefore, the Smarr relation for the mentioned black hole in 4D is obtained from the following relation using the holographic dictionary,
\begin{equation}\label{eq377}
\begin{split}
E=\tilde{T} S+\tilde{\Phi} \tilde{Q}+\tilde{\mathcal{A}} \tilde{\beta}+\mu C
\end{split}
\end{equation}
Also, we have,
\begin{equation}\label{eq51}
\begin{split}
&  d=5\qquad \rightarrow \qquad \delta\bigg( \frac{M}{\omega}\bigg)= \frac{T}{\omega} \delta \bigg(\frac{A}{4G} \bigg)+\bigg(\frac{M}{\omega}-\frac{TS}{\omega}-\frac{Q\Phi}{\omega}-\frac{A\beta}{\omega} \bigg) \frac{\delta (k \ell^{3}/G)}{k \ell^{3}/G}\\
&-\frac{M}{3\omega}\frac{\delta (kR^{3})}{kR^{3}}+\frac{\Phi \sqrt{G}}{\omega\ell} \delta \bigg(\frac{Q\ell}{\sqrt{G}} \bigg)+\frac{A}{\omega \ell}  \delta( \ell\beta),
\end{split}
\end{equation}
or simply as
\begin{equation}\label{eq52}
\begin{split}
&  d=5\qquad \rightarrow \qquad \delta E= \tilde{T} \delta \tilde{S}+\mu \delta C-p \delta \mathcal{V}+ \tilde{\Phi} \delta \tilde{Q}+ \tilde{A} \delta\tilde{\beta}\\
& E=\frac{M}{\omega}, \qquad  \tilde{S}=S,\qquad \tilde{T}=\frac{T}{\omega}, \qquad  \tilde{\Phi}=\frac{\Phi\sqrt{G}}{\omega\ell},\qquad \tilde{Q}=\frac{Q\ell}{\sqrt{G}}, \qquad \tilde{\mathcal{A}}=\frac{\mathcal{A}}{\omega\ell},\qquad \tilde{\beta}=\ell\beta,
\end{split}
\end{equation}
where, using equations \eqref{eq51}, and \eqref{eq52}, we will have,
\begin{equation}\label{eq53}
\begin{split}
&  \mu=\frac{1}{C}(E-\tilde{T}S-\tilde{\Phi}\tilde{Q}-\tilde{\mathcal{A}}\tilde{\beta})\\
&p=\frac{E}{3\mathcal{V}} , \qquad \mathcal{V}=k R^3
\end{split}
\end{equation}
Smarr relation for the this black hole in 5D is as follows,
\begin{equation}\label{eq377}
\begin{split}
E=\tilde{T} S+\tilde{\Phi} \tilde{Q}+\tilde{\mathcal{A}} \tilde{\beta}+\mu C
\end{split}
\end{equation}
Therefore, equation \eqref{eq52} expresses the dictionary between bulk and boundary quantities. Also, by comparing equations \eqref{eq49} and \eqref{eq52}, we find that the dimensions only affect the dictionary, GB parameters, and their conjugates. In the following, we try to obtain the internal energy formula for the GB black hole in the dimensions $d=4$ and $d=5$ for the CFT.\\\\
\\
$\bullet\mathbf{d=4}$
\\\\
As mentioned above, we defined $\beta=\alpha$ for $d=4$. Now, to obtain the internal energy formula for the CFT, we define two dimensionless parameters as follows,
\begin{equation}\label{eq54}
\begin{split}
x\equiv \frac{r_+}{\ell} \qquad \qquad  y\equiv \frac{\tilde{\beta}}{\ell^2}.
\end{split}
\end{equation}
Therefore, according to equations \eqref{eq10}, \eqref{eq12},\eqref{eq23}, \eqref{eq49}, and \eqref{eq54}, we have,
\begin{equation}\label{eq55}
\begin{split}
&E=\frac{1}{32 \pi  C  x \sqrt{k \mathcal{V}}}\bigg(k^2 \tilde{Q}^2+256 \pi ^2 C^2 x^4+256 \pi ^2 C^2 x^2+256 \pi ^2 C^2 y\bigg)\\
&\tilde{S}=\frac{16 \pi ^2 C}{k} \left(x^2+4 y \log \left[\frac{x}{\sqrt{y}}\right]\right).
\end{split}
\end{equation}
We can also calculate thermodynamic quantities for the CFT,
\begin{equation}\label{eq56}
\begin{split}
&\tilde{T}=\bigg(\frac{\partial E}{ \partial \tilde{S}} \bigg)_{\tilde{Q},\mathcal{V},C,\tilde{\beta}}=\sqrt{\frac{k}{\mathcal{V}}}\left(\frac{ 768 \pi ^2 C^2 x^4+256 \pi ^2 C^2 x^2-256 \pi ^2 C^2 y-k^2 \tilde{Q}^2}{1024 \pi ^3 C^2 x \left(x^2+2 y\right)}\right),
\end{split}
\end{equation}
\begin{equation}\label{eq57}
\begin{split}
\tilde{\Phi}=\bigg(\frac{\partial E}{ \partial \tilde{Q}} \bigg)_{\tilde{S},\mathcal{V},C,\tilde{\beta}}=\sqrt{\frac{k^3}{\mathcal{V}}}\bigg(\frac{\tilde{Q}}{16 \pi  C x} \bigg),
\end{split}
\end{equation}
\begin{equation}\label{eq58}
\begin{split}
p=-\bigg(\frac{\partial E}{ \partial \mathcal{V}} \bigg)_{\tilde{S},\tilde{Q},C,\tilde{\beta}}=\frac{E}{2\mathcal{V}},
\end{split}
\end{equation}
\begin{equation}\label{eq59}
\begin{split}
\mu=\bigg(\frac{\partial E}{ \partial C} \bigg)_{\tilde{S},\tilde{Q},\mathcal{V},\tilde{\beta}}=\frac{256 \pi ^2 C^2 x^4+256 \pi ^2 C^2 x^2+256 \pi ^2 C^2 y-k^2 \tilde{Q}^2}{32 \pi  C^2 x \sqrt{k \mathcal{V}}},
\end{split}
\end{equation}
\begin{equation}\label{eq60}
\begin{split}
\tilde{\mathcal{A}}=\bigg(\frac{\partial E}{ \partial \tilde{\beta}} \bigg)_{\tilde{S},\tilde{Q},\mathcal{V},C}=\frac{8 \pi  C y}{x\tilde{\beta} \sqrt{k \mathcal{V}}}.
\end{split}
\end{equation}
In the following, we examine the Gauss-Bonnet black hole for $d=5$.
\\\\
$\bullet\mathbf{d=5}$
\\\\
As mentioned above, we defined $\beta=(d-4)(d-3)\alpha$ for $d=5$. Now, to obtain the internal energy formula for the CFT, we define two dimensionless parameters as follows,
\begin{equation}\label{eq61}
\begin{split}
x\equiv \frac{r_+}{\ell} \qquad \qquad  y\equiv \frac{\tilde{\beta}}{\ell^3},
\end{split}
\end{equation}
Therefore, according to equations \eqref{eq5}, \eqref{eq7},\eqref{eq23}, \eqref{eq52}, and \eqref{eq61}, we have,
\begin{equation}\label{eq62}
\begin{split}
&E=\frac{\Sigma  \left(k^2 \tilde{Q}^2+768 \pi ^2 C^2 x^6+768 \pi ^2 C^2 x^4+768 \pi ^2 C^2 x^2 y\right)}{256 \pi ^2 C (k^{2} \mathcal{V})^{\frac{1}{3}} x^2}\\
&\tilde{S}=\frac{4 \pi  C \Sigma  \left(x^3+6 x y\right)}{k}.
\end{split}
\end{equation}
Also, we can  calculate thermodynamic quantities for the CFT,
\begin{equation}\label{eq63}
\begin{split}
&\tilde{T}=\bigg(\frac{\partial E}{ \partial \tilde{S}} \bigg)_{\tilde{Q},\mathcal{V},C,\tilde{\beta}}= \bigg(\frac{k}{\mathcal{V}}\bigg)^{\frac{1}{3}}\bigg(\frac{-k^2 \tilde{Q}^2+1536 \pi ^2 C^2 x^6+768 \pi ^2 C^2 x^4}{1536 \pi ^3 C^2 x^3 \left(x^2+2 y\right)}\bigg),
\end{split}
\end{equation}
\begin{equation}\label{eq64}
\begin{split}
\tilde{\Phi}=\bigg(\frac{\partial E}{ \partial \tilde{Q}} \bigg)_{\tilde{S},\mathcal{V},C,\tilde{\beta}}=\bigg(\frac{k^4}{\mathcal{V}}\bigg)^{\frac{1}{3}}\bigg(\frac{\Sigma  \tilde{Q}}{128 \pi ^2 c x^2} \bigg),
\end{split}
\end{equation}
\begin{equation}\label{eq65}
\begin{split}
p=-\bigg(\frac{\partial E}{ \partial \mathcal{V}} \bigg)_{\tilde{S},\tilde{Q},C,\tilde{\beta}}=\frac{E}{2\mathcal{V}},
\end{split}
\end{equation}
\begin{equation}\label{eq66}
\begin{split}
\mu=\bigg(\frac{\partial E}{ \partial C} \bigg)_{\tilde{S},\tilde{Q},\mathcal{V},\tilde{\beta}}=\frac{\Sigma  \left(-k^2 \tilde{Q}^2+768 \pi ^2 C^2 x^6+768 \pi ^2 C^2 x^4+768 \pi ^2 C^2 x^2 y\right)}{256 \pi ^2 C^2 k^{2/3} \sqrt[3]{\mathcal{V}} x^2},
\end{split}
\end{equation}
\begin{equation}\label{eq66a}
\begin{split}
\tilde{\mathcal{A}}=\bigg(\frac{\partial E}{ \partial \tilde{\beta}} \bigg)_{\tilde{S},\tilde{Q},\mathcal{V},C}=\frac{3 C \Sigma  y}{\tilde{\beta} \sqrt[3]{k^{2}\mathcal{V}}}.
\end{split}
\end{equation}
Here, $\mu$ is the conjugate of the chemical potential at the boundary, and \eqref{eq59},\eqref{eq65}  represent the equation of state in CFT. The idea of a variable Newton's constant is only relevant in the context of the mixed first law. This law modifies the volume term in the formula for black hole phase transitions. However, the internal energy formula, boundary first law, and Euler relation do not require a dynamic Newton's constant. The focus is on creating an exact dual relation between bulk and boundary first laws. A dynamic parameter called $\omega$ is used to prevent the degeneracy of volume $\mathcal{V}$ and central charge $C$. We will use these variables to examine the different stages in the different thermodynamic ensembles in the dual CFT.
\section{Discussion and results}
In \cite{30',31'}, the thermodynamic phase structure of charged and rotating black holes in the CFT state has been investigated. Now, we are studying the GB AdS black hole at $d=4,5$, which preserves the CFT state. It is important to note that for the GB AdS black hole, at $d=4,5$ the bulk-boundary state of the phase structure remains the same. To determine the phase structure of the system using the canonical ensemble, we need to calculate the Helmholtz energy of the system first. In the following, we will investigate the thermodynamic phase structure of the black hole in dimensions $d=4$ and $d=5$ separately.
\subsection{$d=5$}
\begin{figure}[h!]
 \begin{center}
 \includegraphics[height=6cm,width=7cm]{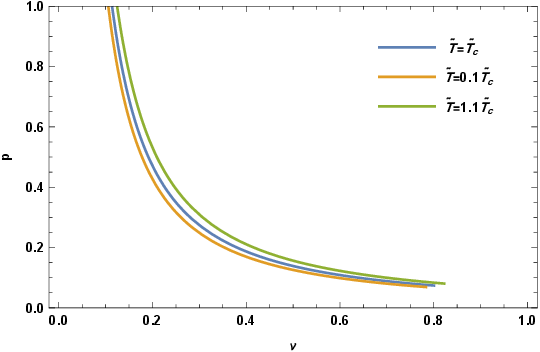}
 \caption{\small{ $p - \mathcal{V}$ criticality for holographic CFT. With the $d=5$, $k=\Sigma=\tilde{Q}=C=1$ and $y=0.01.$}}
 \label{fig1}
 \end{center}
 \end{figure}

\begin{figure}[h!]
 \begin{center}
 \subfigure[]{
 \includegraphics[height=4cm,width=5cm]{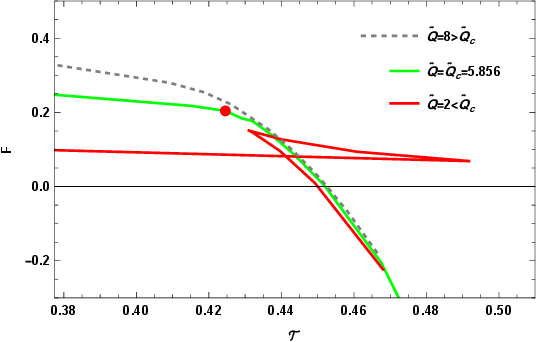}
 \label{fig2a}}
 \subfigure[]{
 \includegraphics[height=4cm,width=5cm]{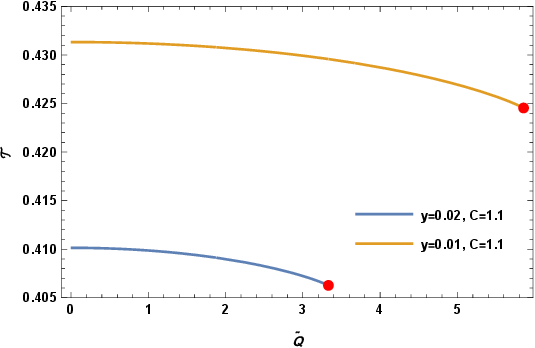}
 \label{fig2b}}
 \subfigure[]{
 \includegraphics[height=4cm,width=5cm]{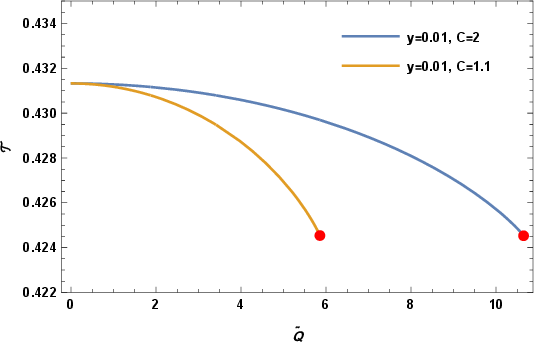}
 \label{fig2c}}
 \caption{\small{With the $k=\Sigma=\mathcal{V}=1$. (a) Free energy $F$ vs. temperature $T$ plot in $d = 5$  with the critical charge $\tilde{Q}_c = 5.856$ for the fixed $(\tilde{Q}, \mathcal{V}, C, y)$ ensemble. The red point is the critical point. With $y=0.01$, and $C=1.1$. (b) The coexistence lines show how the coexistence temperature changes as the charge varies while $y$ is variable and $C$ is fixed. These lines terminate at critical points marked in red. (c) The coexistence lines show how the coexistence temperature changes as the charge varies while $C$ is variable and $y$ is fixed.}}
 \label{fig2}
 \end{center}
 \end{figure}

\begin{figure}[h!]
 \begin{center}
 \subfigure[]{
 \includegraphics[height=4cm,width=5cm]{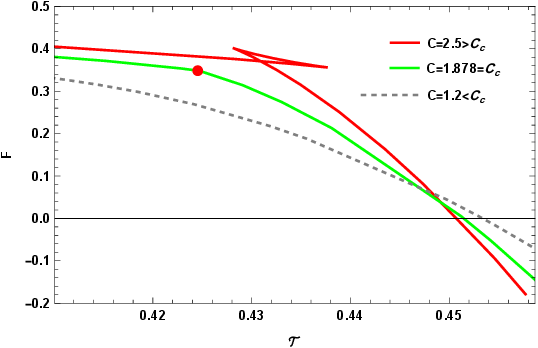}
 \label{fig3a}}
 \subfigure[]{
 \includegraphics[height=4cm,width=5cm]{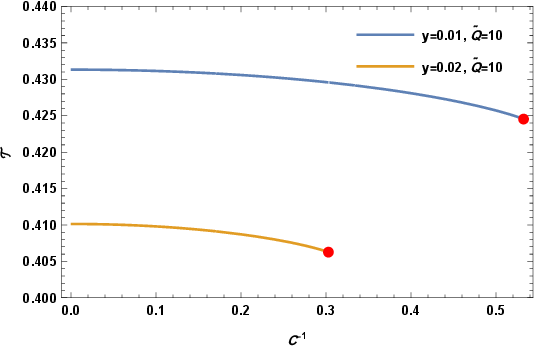}
 \label{fig3b}}
 \subfigure[]{
 \includegraphics[height=4cm,width=5cm]{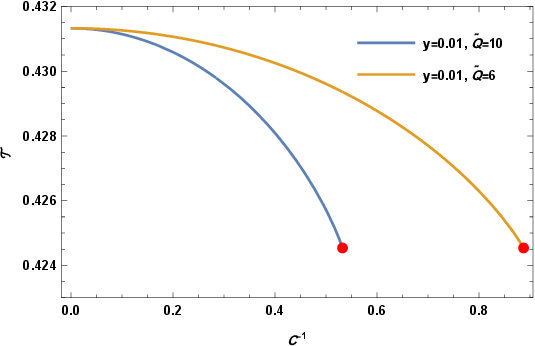}
 \label{fig3c}}
 \caption{\small{With the $k=\Sigma=\mathcal{V}=1$. (a) Free energy $F$ vs. temperature $T$ plot in $d = 5$  with the critical centeral charge $\tilde{C}_c = 1.878$ for the fixed $(\tilde{Q}, \mathcal{V}, C, y)$ ensemble. The red point is the critical point. With $y=0.01$ and $\tilde{Q}=10$. (b) The coexistence lines show how the coexistence temperature changes as the  inverse central charge varies while $y$ is variable and $\tilde{Q}$ is fixed. These lines terminate at critical points marked in red. (c) The coexistence lines show how the coexistence temperature changes as the inverse central varies while $\tilde{Q}$ is variable and $y$ is fixed.}}
 \label{fig3}
 \end{center}
 \end{figure}

We obtain the Helmholtz energy using equations \eqref{eq55} and \eqref{eq56} as follows,
\begin{equation}\label{eq67}
\begin{split}
&F=E-\tilde{T}\tilde{S}=\frac{k^{4/3} \Sigma  \tilde{Q}^2}{256 \pi ^2 C \sqrt[3]{\mathcal{V}} x^2}+\frac{k \Sigma  \tilde{Q}^2 \left(x^2+6 y\right)}{384 \pi ^2 C x^2 \sqrt[3]{\frac{\mathcal{V}}{k}} \left(x^2+2 y\right)}\\
&+\frac{3 C \Sigma  \left(x^4+x^2+y\right)}{k^{2/3} \sqrt[3]{\mathcal{V}}}-\frac{2 C \Sigma  x^2 \left(2 x^2+1\right) \left(\frac{\mathcal{V}}{k}\right)^{2/3} \left(x^2+6 y\right)}{\mathcal{V} \left(x^2+2 y\right)}.
\end{split}
\end{equation}
Also, we can use the following relationships to determine the critical points,
\begin{equation}\label{eq68}
\begin{split}
\bigg(\frac{\partial \tilde{T}}{ \partial x} \bigg)_{\tilde{Q},\mathcal{V},C,\tilde{\beta}}=0, \qquad   \bigg(\frac{\partial^2 \tilde{T}}{ \partial x^2} \bigg)_{\tilde{Q},\mathcal{V},C,\tilde{\beta}}=0.
\end{split}
\end{equation}
Therefore, by using the relations between \eqref{eq56} and \eqref{eq68}, we can find the critical points as follows,
\begin{equation}\label{eq69}
\begin{split}
x_c=\frac{\sqrt{\sqrt{2916 y^2-900 y+25}-54 y+5}}{\sqrt{30}},
\end{split}
\end{equation}
\begin{equation}\label{eq70}
\begin{split}
\tilde{Q}_c^2=\frac{768 \pi ^2 C^2 x_c^4 \left(-12 y x_c^2-2 x_c^4+x_c^2-2 y\right)}{k^2 \left(5 x_c^2+6 y\right)},\quad \text{or}   \quad   C_c^2=\frac{k^2 \tilde{Q}^2 \left(5 x_c^2+6 y\right)}{768 \pi ^2 x_c^4 \left(-12 y x_c^2-2 x_c^4+x_c^2-2 y\right)},
\end{split}
\end{equation}
and
\begin{equation}\label{eq71}
\begin{split}
&\tilde{T_c}=\bigg({\frac{k}{\mathcal{V}}}\bigg)^{\frac{1}{3}}\frac{2x_c \big(3 x_c^2+1\big)}{\pi  \big(5 x_c^2+6 y\big)},\quad \tilde{\Phi}_c=\frac{k^{4/3} \Sigma }{128 \pi ^2 \sqrt[3]{\mathcal{V}} x_c^2}
\sqrt{\frac{768 \pi ^2 x_c^4 \left(-12 y x_c^2-2 x_c^4+x_c^2-2 y\right)}{k^2 \left(5 x_c^2+6 y\right)}}\\
&\mu_c=\frac{3 \Sigma  \left(7 x_c^6+18 x_c^4 y+4 x_c^4+13 x_c^2 y+6 y^2\right)}{k^{2/3} \sqrt[3]{\mathcal{V}} \left(5 x_c^2+6 y\right)}, \\
&p_c=\frac{9 \Sigma  \left(2x_c^4-2 y x_c^4+3 y x_c^2+x_c^6+2 y^2\right)}{2 k^{2/3} \mathcal{V}^{4/3} \left(5 x_c^2+6 y\right)}\sqrt{\frac{k^2 \tilde{Q} \left(5 x_c^2+6 y\right)}{768 \pi ^2 x_c^4 \left(x_c^2-12 y x_c^2-2 x_c^4-2 y\right)}}\\
&\mathcal{A}_c=\frac{3  \Sigma  y}{\tilde{\beta} \sqrt[3]{k^{2}\mathcal{V}}}\sqrt{\frac{k^2 \tilde{Q} \left(5 x_c^2+6 y\right)}{768 \pi ^2 x_c^4 \left(x_c^2-12 y x_c^2-2 x_c^4-2 y\right)}}.
\end{split}
\end{equation}
According to the above relations, only $x_c$ and $C_c$ are independent of the volume of the CFT and only depend on the GB parameter and the charge of the black hole. Also, the critical temperature tends to zero only when $\mathcal{V}\rightarrow \infty$. Note that we will only have a physical $x_c$ (corresponding to the critical event horizon) in the situation where $y<0.031$. It is very interesting that when we turn off the Gauss-Bonnet parameter, the results obtained for the critical points are in agreement with \cite{30'}. Similar to the case of charged and angular momentum AdS black holes discussed in \cite{30',31'}, we observe that the CFT states corresponding to GB AdS black holes cannot be compared directly to Van der Waals fluids. Because, according to relations \eqref{eq62} and \eqref{eq65}, we have $p\propto \mathcal{V}^{\frac{4}{3}}$, you can also see this inconsistency in Figure \eqref{fig1}.
We use the temperature \eqref{eq56} and free energy \eqref{eq67} to plot Figure \eqref{fig2}. In this figure, we depict the variations in free energy for various charge values and display the coexistence curves for different central charge  and $y$ (corresponding to the GB parameter) values. In Figure \eqref{fig2a}, we just keep $C$, $\mathcal{V}$, and $y$ fixed.
Also, the free energy shows a smooth monotonic curve for $\tilde{Q} > \tilde{Q}_c$, a “swallowtail” shape for $\tilde{Q} <\tilde{Q}_c$, and a kink when $\tilde{Q}=\tilde{Q}_c$.
Each curve starts from the point where $T = 0$ and as $T$ increases, the value of $x$ along the curves also increases. According to the formula $\tilde{S}=\frac{4 \pi  C \Sigma  \left(x^3+6 x y\right)}{k}$ for CFT entropy, it can be observed that black holes with small $x = \frac{r_h}{\ell}$ correspond to CFT thermal states with small $S/C$. In other words, these states have low entropy per degree of freedom. When $\tilde{Q}<\tilde{Q}_c$, the free energy curve exhibits a "swallowtail" shape, indicating a first-order phase transition at a self-intersection point in the CFT. This transition involves a shift from a low entropy per degree of freedom state to a high entropy per degree of freedom state. If $\tilde{Q}>\tilde{Q}_c$, it is not possible to observe distinct phases in the CFT states. Also, We can  find that there is a critical point in the free energy curve when $\tilde{Q}=\tilde{Q}_c$. In this situation, the CFT state experiences a second-order phase transition, where the entropy $S$ of the CFT does not exhibit any sudden change when transitioning from a low entropy per degree of freedom state to a higher entropy per degree of freedom state. In Figure \eqref{fig2b}, we show the coexistence line of CFT modes for different values of $y$. As we know, this coexistence line represents the first-order phase transition between low-entropy (below the coexistence line) and high-entropy states (above the coexistence line). Also, this line ends at a critical point where $\tilde{Q}=\tilde{Q}_c$, indicating a second-order phase transition. From Figure \eqref{fig2b}, we can see that as the value of $y$ increases (C fixed), the coexistence line becomes smaller. As a result, the range of the first-order phase transition also decreases. We also find that the coexistence temperature $(\mathcal{T})$ decreases with increasing charge $\tilde{Q}$.  In Figure \eqref{fig2c}, we show the coexistence line of CFT modes for different values of $C$. In Figure \eqref{fig2b}, we can see that the coexistence temperature decreases as the charge increases. Also, we find from Figure \eqref{fig2c} that with the increase in central charge, the coexistence line becomes larger, thus expanding the range of the first-order phase transition. By comparing figures \eqref{fig2b} and \eqref{fig2c}, we realize that with variable $y$ and fixed $C$, we have different temperatures $\mathcal{T}$ at the point $\tilde{Q}=0$. But for variable $C$ and fixed $y$, we have the same temperature at the point $\tilde{Q}=0$. Therefore, GB parameter can play an important role in the coexistence diagram.
Figure \eqref{fig3} displays the free energy's behavior to various central charges, namely when $C=C_c, C>C_c, C<C_c$. It can be observed that the free energy exhibits similar kink behaviors, swallowtail, and smooth monotonic curve  as those depicted in Figure \eqref{fig2}. As you can see in Figure \eqref{fig3}, a first-order phase transition and a swallowtail shape occur when $C>C_c$, whereas in Figure \eqref{fig2}, this behavior occurs when $\tilde{Q}<\tilde{Q}_c$. Also, the obtained results are in agreement with \cite{30',31',32'}.
Furthermore, if we graph the coexistence curve that illustrates the change in the coexistence central charge in relation to temperature, we will observe that the coexistence temperature decreases as the central charge decreases. In addition, according to Figures \eqref{fig3b} and \eqref{fig3c}, we find that the coexistence temperature increases with the increase of the central charge, and this is in contrast to the relationship between the coexistence temperature and charge in Figure \eqref{fig2}.
Also, based on the coexistence diagrams in Figure \eqref{fig3}, we observe that as $y$ increases, $\tilde{Q}$ remains fixed and  $\tilde{Q}$ increases, $y$ remains fixed, resulting in a shorter coexistence line.

\subsection{$d=4$}

\begin{figure}[h!]
 \begin{center}
 \subfigure[]{
 \includegraphics[height=4cm,width=5cm]{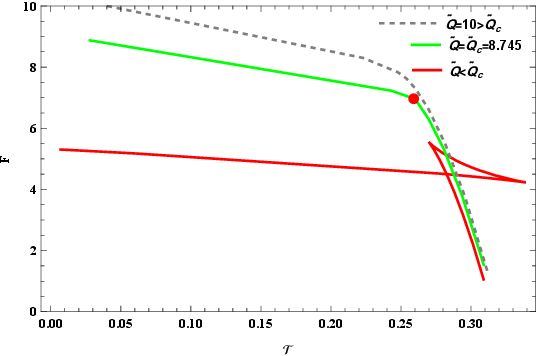}
 \label{fig4a}}
 \subfigure[]{
 \includegraphics[height=4cm,width=5cm]{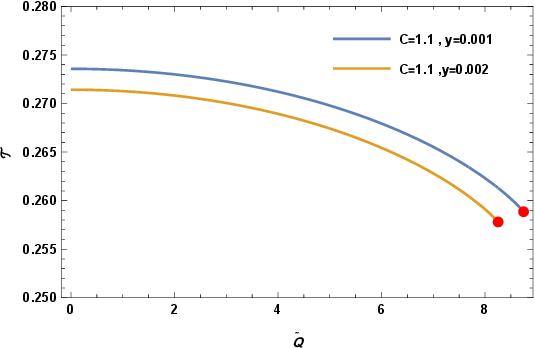}
 \label{fig4b}}
 \subfigure[]{
 \includegraphics[height=4cm,width=5cm]{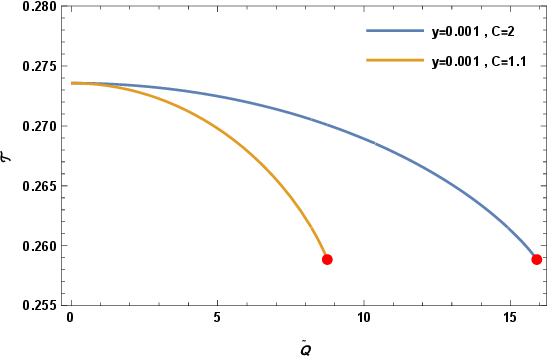}
 \label{fig4c}}
 \caption{\small{With the $k=\Sigma=\mathcal{V}=1$. (a) Free energy $F$ vs. temperature $T$ plot in $d = 4$  with the critical charge $\tilde{Q}_c = 8.745$ for the fixed $(\tilde{Q}, \mathcal{V}, C, y)$ ensemble. The red point is the critical point. With $y=0.001$ and $C=1.1$. (b) The coexistence lines show how the coexistence temperature changes as the charge varies while $y$ is variable and $C$ is fixed. These lines terminate at critical points marked in red. (c) The coexistence lines show how the coexistence temperature changes as the charge varies while $C$ is variable and $y$ is fixed.}}
 \label{fig4}
 \end{center}
 \end{figure}

We use equations \eqref{eq55} and \eqref{eq56} to obtain the Helmholtz energy,
\begin{equation}\label{eq72}
\begin{split}
&F=E-\tilde{T}\tilde{S}= \frac{1}{64 \pi  C  x \left(x^2+2 y\right)\sqrt{\mathcal{V}k}} \bigg[ -256 \pi ^2 C^2 \left(x^6-x^4 (4 y+1)-7 x^2 y-4 y^2\right)  \\
& +k^2 \tilde{Q}^2 \left(3 x^2+4 y\right)+4 y \log \left(\frac{x}{\sqrt{y}}\right) \left(k^2 \tilde{Q}^2-256 \pi ^2 C^2 \left(3 x^4+x^2-y\right)\right)   \bigg]
\end{split}
\end{equation}
Now, using equation \eqref{eq68}, we calculate the critical points for thermodynamic quantities,
\begin{equation}\label{eq73}
\begin{split}
x_c=\frac{\sqrt{\sqrt{144 y^2-40 y+1}-12 y+1}}{2 \sqrt{3}},
\end{split}
\end{equation}
\begin{equation}\label{eq74}
\begin{split}
&\tilde{Q}_c^2=\frac{256 \pi ^2 C^2 \left(-18 y x_c^4-5 y x_c^2-3 x_c^6+x_c^4-2 y^2\right)}{k^2 \left(3 x_c^2+2 y\right)},\\
&C_c^2=\frac{k^2 \tilde{Q}^2 \left(3 x_c^2+2 y\right)}{256 \pi ^2 \left(-18 y x_c^4-5 y x_c^2-3 x_c^6+x_c^4-2 y^2\right)},
\end{split}
\end{equation}
and
\begin{equation}\label{eq75}
\begin{split}
&\tilde{T_c}= \sqrt{\frac{k}{\mathcal{V}}} \frac{x_c \left(6 x_c^2+1\right)}{2 \pi  \left(3 x_c^2+2 y\right)}  ,\quad \tilde{\Phi}_c=\frac{k^{3/2}}{16 \pi   x_c\sqrt{\mathcal{V}} }\sqrt{\frac{256 \pi ^2  \left(-18 y x_c^4-5 y x_c^2-3 x_c^6+x_c^4-2 y^2\right)}{k^2 \left(3 x_c^2+2 y\right)}}     \\
&\mu_c=\frac{16 \pi  \left(x_c^4+10yx_c^4+5 y x_c^2+3 x_c^6+2 y^2\right)}{\sqrt{k \mathcal{V}} x_c \left(3 x_c^2+2 y\right)}, \\
&p_c= \frac{\sqrt{k} \tilde{Q} (1-4 y) x_c^3}{\mathcal{V}^{3/2} \left(3 x_c^2+2 y\right)} \sqrt{\frac{3 x_c^2+2 y}{x_c^4-18 y x_c^4-5 y x_c^2-3 x_c^6-2 y^2}},\\
&\mathcal{A}_c=\sqrt{\frac{k}{\mathcal{V}}}\frac{y\tilde{Q}(3 x_c^2+2 y)}{2x_c \tilde{\beta}(-18 y x_c^4-5 y x_c^2-3 x_c^6+x_c^4-2 y^2)}.
\end{split}
\end{equation}
Note that we will only have a physical $x_c$ (corresponding to the critical event horizon) in the situation where $y<0.028$. The critical points of thermodynamic quantities for $d=4$ have a behavior similar to that of $d=5$. It is important to note that $x_c$ and $C_c$ are independent of the CFT volume, and the critical temperature only tends to zero if the CFT volume tends to infinity. According to relations \eqref{eq55} and \eqref{eq58}, we have $p\propto \mathcal{V}^{\frac{3}{2}}$, so we cannot expect the same behavior as van der Waals fluids \cite{31'}.
Next, we use equations  \eqref{eq56} and  \eqref{eq72} to draw the Helmholtz energy diagram in terms of temperature. Also, we draw the coexistence lines for different conditions in $d=4$.
According to Figure \eqref{fig4}, we find that the behavior of phase transition in CFT states, for $d=4$ is completely similar to $d=5$ and they differ only in the values of the critical points.
Also, in the bulk-boundary case, there is a similar behavior for the GB AdS black hole in $D=4$ and $D=5$ \cite{33'}.
\newpage
\section{Conclusions}
In this paper, we have studied the thermodynamics and phase transition of charged Gauss-Bonnet black holes in anti-de Sitter (AdS) space, using a holographic approach that treats the cosmological constant $\Lambda$ and the Newton constant $G$ as thermodynamic variables. The cosmological constant $\Lambda$ can be interpreted as the thermodynamic pressure $P$ of the black hole, and its conjugate quantity is the thermodynamic volume $V$ of the black hole. The Newton constant $G$ can be related to the central charge $C$ of the dual conformal field theory (CFT) that lives on the boundary of the AdS space, according to the AdS/CFT correspondence. We have shown that the central charge $C$ of the dual CFT is related to the variation of $\Lambda$ and $G$, and that it can be kept constant by adjusting them simultaneously. This means that we can study the thermodynamics and phase transition of the black hole without changing the properties of the dual CFT. We have derived the modified first law of thermodynamics that includes the thermodynamic volume $V$ and pressure $P$, as well as the electric charge $Q$ and potential $\Phi$ of the black hole. We have investigated the phase structures of the black holes by using the free energy $F$, and found that there exists a critical value of the central charge $C_c$, above which the phase diagram exhibits a first-order phase transition between small and large black holes. This phase transition is similar to the liquid-gas phase transition of a van der Waals fluid, and it is characterized by the critical exponents and the coexistence curve. We have also discussed the implications of our results for the holographic complexity $C$ of the CFT. We have found that both quantities exhibit non-monotonic behaviors near the phase transition and that they can be used to distinguish different phases of the black hole. We have concluded that the central charge $C$ plays an important role in the holographic thermodynamics and phase transition of charged gauss-bonnet black holes, and that it can be used as a probe of the quantum gravity effects in the bulk. Our work opens up new avenues for exploring the connection between black hole chemistry and holography, and for understanding the nature of quantum gravity in higher dimensions.  Also, we conclude that although the central charge we considered has suitable and satisfactory solutions for this black hole in 4 and 5 dimensions, it is necessary to introduce a special central charge for this structure of modified gravity. In other words, the central charge in holographic thermodynamics is not a universal value and needs to be modified according to different modified gravities. Therefore, the physics of the problem will be very different from the one discussed in this article.


\begin{thebibliography}{11}
\bibitem{10000}
Bardeen, James M., Brandon Carter, and Stephen W. Hawking. "The four laws of black hole mechanics." Communications in mathematical physics 31 (1973): 161-170.
\bibitem{1000}
Davies, Paul CW. "Thermodynamics of black holes." Reports on Progress in Physics 41.8 (1978): 1313.
\bibitem{1001}
Hawking, Stephen W., and Don N. Page. "Thermodynamics of black holes in anti-de Sitter space." Communications in Mathematical Physics 87 (1983): 577-588.
\bibitem{1002}
Wald, Robert M. "The thermodynamics of black holes." Living reviews in relativity 4 (2001): 1-44.
\bibitem{1003}
Hawking, Stephen W. "Black holes and thermodynamics." Physical Review D 13.2 (1976): 191.
\bibitem{1004}
Bekenstein, Jacob D. "Black‐hole thermodynamics." Physics Today 33.1 (1980): 24-31.
\bibitem{1005}
Wald, Robert M. "Black holes and thermodynamics." Black holes and relativistic stars (1998): 155-176.
\bibitem{1006}
Hayward, Sean A. "Unified first law of black-hole dynamics and relativistic thermodynamics." Classical and Quantum Gravity 15.10 (1998): 3147.
\bibitem{1007}
Carter, Benedict MN, and Ishwaree P. Neupane. "Thermodynamics and stability of higher dimensional rotating (Kerr-) AdS black holes." Physical Review D 72.4 (2005): 043534.
\bibitem{1008}
Ruppeiner, George. "Thermodynamic black holes." Entropy 20.6 (2018): 460.
\bibitem{1009}
Xiao, Yong, Yu Tian, and Yu-Xiao Liu. "Extended black hole thermodynamics from extended Iyer-Wald formalism." Physical Review Letters 132.2 (2024): 021401.
\bibitem{1010}
Minamitsuji, Masato, and Kei-ichi Maeda. "Black hole thermodynamics in generalized Proca theories." arXiv preprint arXiv:2403.08986 (2024).
\bibitem{1011}
Bardeen, James M. "Kerr metric black holes." Nature 226.5240 (1970): 64-65.
\bibitem{1011'}
Bardeen, James M. "Properties of black holes relevant to their observation." Symposium-International Astronomical Union. Vol. 64. Cambridge University Press, 1974.
\bibitem{1011''}
Bardeen, James M. "Timelike and null geodesics in the Kerr metric." Black holes 215 (1973).
\bibitem{1}
Kastor, David, Sourya Ray, and Jennie Traschen. "Enthalpy and the mechanics of AdS black holes." Classical and Quantum Gravity 26.19 (2009): 195011.
\bibitem{2}
Kubiznak, David, and Robert B. Mann. "P-V criticality of charged AdS black holes." Journal of High Energy Physics 2012.7 (2012): 1-25.
\bibitem{3}
Cai, Rong-Gen. "Gauss-Bonnet black holes in AdS spaces." Physical Review D 65.8 (2002): 084014.
\bibitem{4}
Altamirano, Natacha, et al. "Kerr-AdS analogue of triple point and solid/liquid/gas phase transition." Classical and Quantum Gravity 31.4 (2014): 042001.
\bibitem{5}
Altamirano, Natacha, David Kubizňák, and Robert B. Mann. "Reentrant phase transitions in rotating anti–de Sitter black holes." Physical Review D 88.10 (2013): 101502.
\bibitem{6}
Dutta, Suvankar, Akash Jain, and Rahul Soni. "Dyonic black hole and holography." Journal of High Energy Physics 2013.12 (2013): 1-30.
\bibitem{7}
Johnson, Clifford V. "Holographic heat engines." Classical and Quantum Gravity 31.20 (2014): 205002.
\bibitem{8}
Kubizňák, David, and Robert B. Mann. "Black hole chemistry." Canadian Journal of Physics 93.9 (2015): 999-1002.
\bibitem{9}
Kubizňák, David, Robert B. Mann, and Mae Teo. "Black hole chemistry: thermodynamics with Lambda." Classical and Quantum Gravity 34.6 (2017): 063001.
\bibitem{1013}
Banerjee, Rabin, and Dibakar Roychowdhury. "Thermodynamics of phase transition in higher dimensional AdS black holes." Journal of High Energy Physics 2011.11 (2011): 1-13.
\bibitem{1014}
Mandal, Abhijit, Saurav Samanta, and Bibhas Ranjan Majhi. "Phase transition and critical phenomena of black holes: A general approach." Physical Review D 94.6 (2016): 064069.
\bibitem{1015}
Maity, Reevu, Pratim Roy, and Tapobrata Sarkar. "Black hole phase transitions and the chemical potential." Physics Letters B 765 (2017): 386-394.
\bibitem{1016}
Silva, Pedro J. "Phase transitions and statistical mechanics for BPS black holes in AdS/CFT." Journal of High Energy Physics 2007.03 (2007): 015.
\bibitem{1017}
El Moumni, H. "Revisiting the phase transition of AdS-Maxwell–power-Yang–Mills black holes via AdS/CFT tools." Physics Letters B 776 (2018): 124-132.
\bibitem{1018}
Sokolowski, Leszek M., and Pawel Mazur. "Second-order phase transitions in black-hole thermodynamics." Journal of Physics A: Mathematical and General 13.3 (1980): 1113.
\bibitem{1019}
Wei, Shao-Wen, and Yu-Xiao Liu. "Insight into the microscopic structure of an AdS black hole from a thermodynamical phase transition." Physical review letters 115.11 (2015): 111302.
\bibitem{1020}
Li, Ran, and Jin Wang. "Thermodynamics and kinetics of Hawking-Page phase transition." Physical Review D 102.2 (2020): 024085.
\bibitem{1021}
Aharony, Ofer, Erez Y. Urbach, and Maya Weiss. "Generalized Hawking-Page transitions." Journal of High Energy Physics 2019.8 (2019): 1-21.
\bibitem{1022}
Wei, Shao-Wen, Yu-Xiao Liu, and Robert B. Mann. "Novel dual relation and constant in Hawking-Page phase transitions." Physical Review D 102.10 (2020): 104011.
\bibitem{1023}
Belhaj, Adil, et al. "On universal constants of AdS black holes from Hawking-Page phase transition." Physics Letters B 811 (2020): 135871.
\bibitem{1024}
Barzi, F., H. El Moumni, and K. Masmar. "Rényi topology of charged-flat black hole: Hawking-Page and Van-der-Waals phase transitions." Journal of High Energy Astrophysics (2024).
\bibitem{a}
Wei, Shao-Wen, and Yu-Xiao Liu. "Topology of black hole thermodynamics." Physical Review D 105.10 (2022): 104003.
\bibitem{b}
Yerra, Pavan Kumar, and Chandrasekhar Bhamidipati. "Topology of black hole thermodynamics in Gauss-Bonnet gravity." Physical Review D 105.10 (2022): 104053.
\bibitem{c}
Bai, Ning-Chen, Lei Li, and Jun Tao. "Topology of black hole thermodynamics in Lovelock gravity." Physical Review D 107.6 (2023): 064015.
\bibitem{d}
Wei, Shao-Wen, Yu-Xiao Liu, and Robert B. Mann. "Black hole solutions as topological thermodynamic defects." Physical Review Letters 129.19 (2022): 191101.
\bibitem{e}
Wu, Di. "Topological classes of rotating black holes." Physical Review D 107.2 (2023): 024024.
\bibitem{f}
Wu, Di. "Topological classes of thermodynamics of the four-dimensional static accelerating black holes." Physical Review D 108.8 (2023): 084041.
\bibitem{g}
Sadeghi, Jafar, et al. "Thermodynamic topology of black holes from bulk-boundary, extended, and restricted phase space perspectives." Annals of Physics (2023): 169569.
\bibitem{h}
Zhang, Meng-Yao, et al. "Topology of nonlinearly charged black hole chemistry via massive gravity", The European Physical Journal C 83 (2023):773
\bibitem{i}
Alipour, Mohammad Reza, et al. "Topological classification and black hole thermodynamics." Physics of the Dark Universe 42, (2023):101361.
\bibitem{j}
Sadeghi, Jafar, et al. "Bulk-boundary and RPS Thermodynamics from Topology perspective." arXiv preprint arXiv:2306.16117 (2023).
\bibitem{k}
Sadeghi, Jafar, et al. "Bardeen black hole thermodynamics from topological perspective." Annals of Physics (2023): 169391.
\bibitem{l}
Zhang, Ming, and Jie Jiang. "Bulk-boundary thermodynamic equivalence: a topology viewpoint." Journal of High Energy Physics 2023.6 (2023): 1-17.
\bibitem{m}
Sadeghi, Jafar, et al. "Topology of Hayward-AdS black hole thermodynamics." Phys. Scr. 99 (2024): 025003 .
\bibitem{n}
Sadeghi, Jafar, et al. "Thermodynamic topology and photon spheres in the hyperscaling violating black holes." Astroparticle Physics (2023): 102920.
\bibitem{1025}
Dutta, Suvankar, and Gurmeet Singh Punia. "String theory corrections to holographic black hole chemistry." Physical Review D 106.2 (2022): 026003.
\bibitem{1026}
Sinamuli, Musema, and Robert B. Mann. "Higher order corrections to holographic black hole chemistry." Physical Review D 96.8 (2017): 086008.
\bibitem{1027}
Mir, Mozhgan, et al. "Black hole chemistry and holography in generalized quasi-topological gravity." Journal of High Energy Physics 2019.8 (2019): 1-70.
\bibitem{1028}
Kubizňák, David, Robert B. Mann, and Mae Teo. "Black hole chemistry: thermodynamics with Lambda." Classical and Quantum Gravity 34.6 (2017): 063001.
\bibitem{1029}
De Haro, Sebastian, Kostas Skenderis, and Sergey N. Solodukhin. "Holographic reconstruction of spacetime and renormalization in the ads/cft correspondence." Communications in Mathematical Physics 217 (2001): 595-622.
\bibitem{1030}
Rivelles, Victor O. "Holographic principle and ads/cft correspondence." arXiv preprint hep-th/9912139 (1999).
\bibitem{1031}
Dobashi, Suguru, and Tamiaki Yoneya. "Resolving the holography in the plane-wave limit of AdS/CFT correspondence." Nuclear Physics B 711.1-2 (2005): 3-53.
\bibitem{1032}
Damghan, Iran. "Holography and its Applications (ICHA2 2023)." (2023).
\bibitem{10}
Karch, Andreas, and Brandon Robinson. "Holographic black hole chemistry." Journal of High Energy Physics 2015.12 (2015): 1-15.
\bibitem{11}
Mir, Mozhgan, et al. "Black hole chemistry and holography in generalized quasi-topological gravity." Journal of High Energy Physics 2019.8 (2019): 1-70.
\bibitem{12}
Karch, Andreas, and Brandon Robinson. "Holographic black hole chemistry." Journal of High Energy Physics 2015.12 (2015): 1-15.
\bibitem{13}
Sinamuli, Musema, and Robert B. Mann. "Higher order corrections to holographic black hole chemistry." Physical Review D 96.8 (2017): 086008.
\bibitem{14}
Visser, Manus R. "Holographic thermodynamics requires a chemical potential for color." Physical Review D 105.10 (2022): 106014.
\bibitem{15}
Gibbons, Gary, Renata Kallosh, and Barak Kol. "Moduli, scalar charges, and the first law of black hole thermodynamics." Physical review letters 77.25 (1996): 4992.
\bibitem{16}
Creighton, Jolien DE, and Robert B. Mann. "Quasilocal thermodynamics of dilaton gravity coupled to gauge fields." Physical Review D 52.8 (1995): 4569.
\bibitem{17}
Cong, Wan, David Kubizňák, and Robert B. Mann. "Thermodynamics of AdS black holes: critical behavior of the central charge." Physical Review Letters 127.9 (2021): 091301.
\bibitem{18}
Qu, Yang, Jun Tao, and Huan Yang. "Thermodynamics and phase transition in central charge criticality of charged Gauss-Bonnet AdS black holes." Nuclear Physics B 992 (2023): 116234.
\bibitem{1033}
Parikh, Maulik. "Enhanced instability of de Sitter space in Einstein-Gauss-Bonnet gravity." Physical Review D 84.4 (2011): 044048.
\bibitem{1034}
Chakravarti, Kabir, Rajes Ghosh, and Sudipta Sarkar. "Constraining the topological Gauss-Bonnet coupling from GW150914." Physical Review D 106.4 (2022): L041503.
\bibitem{1035}
Glavan, Dražen, and Chunshan Lin. "Einstein-Gauss-Bonnet gravity in four-dimensional spacetime." Physical review letters 124.8 (2020): 081301.
\bibitem{19}
Alfaia, R. B., I. P. Lobo, and L. C. T. Brito. "Central charge criticality of charged AdS black hole surrounded by different fluids." The European Physical Journal Plus 137.3 (2022): 402.
\bibitem{20}
Gao, Zeyuan, and Liu Zhao. "Restricted phase space thermodynamics for AdS black holes via holography." Classical and Quantum Gravity 39.7 (2022): 075019.
\bibitem{21}
Gao, Zeyuan, Xiangqing Kong, and Liu Zhao. "Thermodynamics of Kerr-AdS black holes in the restricted phase space." The European Physical Journal C 82.2 (2022): 112.
\bibitem{21'}
Alipour, Mohammad Reza, Jafar Sadeghi, and Mehdi Shokri. "WGC and WCCC of black holes with quintessence and cloud strings in RPS space." Nuclear Physics B 990 (2023): 116184.
\bibitem{22}
Sadeghi, Jafar, et al. "RPS thermodynamics of Taub–NUT AdS black holes in the presence of central charge and the weak gravity conjecture." General Relativity and Gravitation 54.10 (2022): 129.
\bibitem{23}
Kumar, Neeraj, Soham Sen, and Sunandan Gangopadhyay. "Phase transition structure and breaking of universal nature of central charge criticality in a Born-Infeld AdS black hole." Physical Review D 106.2 (2022): 026005.
\bibitem{24}
Wang, Yi, and Jie Ren. "Thermodynamics of hairy accelerating black holes in gauged supergravity and beyond." Physical Review D 106.10 (2022): 104046.
\bibitem{25}
Lobo, Iarley P., et al. "Holographic dictionary for generic asymptotically AdS black holes." arXiv preprint arXiv:2206.13664 (2022).
\bibitem{26}
Ghosh, Aritra, Chandrasekhar Bhamidipati, and Sudipta Mukherji. "Logarithmic corrections to black hole entropy and holography." arXiv preprint arXiv:2207.02820 (2022).
\bibitem{27}
Bai, Yan-Ying, et al. "Revisit on thermodynamics of BTZ black hole with variable Newton constant." arXiv preprint arXiv:2208.11859 (2022).
\bibitem{28}
Dutta, Suvankar, and Gurmeet Singh Punia. "String theory corrections to holographic black hole chemistry." Physical Review D 106.2 (2022): 026003.
\bibitem{29}
Ahmed, Moaathe Belhaj, et al. "Holographic dual of extended black hole thermodynamics." Physical Review Letters 130.18 (2023): 181401.
\bibitem{30}
Johnson, Clifford V. "Holographic heat engines." Classical and Quantum Gravity 31.20 (2014): 205002.
\bibitem{31}
Dolan, Brian P. "Bose condensation and branes." Journal of High Energy Physics 2014.10 (2014): 1-8.
\bibitem{32}
Kastor, David, Sourya Ray, and Jennie Traschen. "Chemical potential in the first law for holographic entanglement entropy." Journal of High Energy Physics 2014.11 (2014): 1-17.
\bibitem{33}
Zhang, Jia-Lin, Rong-Gen Cai, and Hongwei Yu. "Phase transition and thermodynamical geometry for Schwarzschild AdS black hole in AdS5× S5 spacetime." Journal of High Energy Physics 2015.2 (2015): 1-16.
\bibitem{34}
Dolan, Brian P. "Pressure and compressibility of conformal field theories from the AdS/CFT correspondence." Entropy 18.5 (2016): 169.
\bibitem{35}
Maldacena, Juan. "The large-N limit of superconformal field theories and supergravity." International journal of theoretical physics 38.4 (1999): 1113-1133.
\bibitem{29'}
Visser, Manus R. "Holographic thermodynamics requires a chemical potential for color." Physical Review D 105.10 (2022): 106014.
\bibitem{30'}
Ahmed, Moaathe Belhaj, et al. "Holographic dual of extended black hole thermodynamics." Physical Review Letters 130.18 (2023): 181401.
\bibitem{300'}
Cong, Wan, et al. "Holographic CFT phase transitions and criticality for charged AdS black holes." Journal of High Energy Physics 2022.8 (2022): 1-37.
\bibitem{31'}
Gong, Ting-Feng, Jie Jiang, and Ming Zhang. "Holographic thermodynamics of rotating black holes." Journal of High Energy Physics 2023.6 (2023): 1-22.
\bibitem{32'}
Ahmed, Moaathe Belhaj, et al. "Holographic CFT phase transitions and criticality for rotating AdS black holes." Journal of High Energy Physics 2023.8 (2023): 1-32.
\bibitem{33'}
Qu, Yang, Jun Tao, and Huan Yang. "Thermodynamics and phase transition in central charge criticality of charged Gauss-Bonnet AdS black holes." Nuclear Physics B 992 (2023): 116234.
\bibitem{33''}
Li, Hui-Ling, Xiao-Xiong Zeng, and Rong Lin. "Holographic phase transition from novel Gauss–Bonnet AdS black holes." The European Physical Journal C 80.7 (2020): 652.
\end{thebibliography}
\end{document}